\documentclass[runningheads]{llncs}
\usepackage[utf8]{inputenc}
\usepackage{longtable}

\usepackage{amssymb}  
\usepackage{amsmath} 

\usepackage{graphicx}
\usepackage{epsfig}
\usepackage{epstopdf}
\usepackage{color}
\usepackage{subfig}
\usepackage{float} 
\usepackage{verbatim}

\usepackage{tikz}
\usetikzlibrary{shapes,arrows,calc,positioning}
\tikzstyle{bigblock} = [draw, fill=blue!20, rectangle, 
    minimum height=6em, minimum width=8em]
\tikzstyle{medblock} = [draw, fill=blue!20, rectangle, 
    minimum height=4em, minimum width=4em]    
\tikzstyle{mux} = [draw, fill=black!20, rectangle, 
    minimum height=5em, minimum width=0.1em]    
\tikzstyle{smallblock} = [draw, fill=blue!20, rectangle, 
    minimum height=3em, minimum width=4em]
\tikzstyle{sum} = [draw, fill=blue!20, circle, node distance=1cm]
\tikzstyle{signal} = [coordinate]
\tikzstyle{pinstyle} = [pin edge={to-,thin,black}]
\tikzstyle{block} = [draw, fill=blue!20, rectangle, 
    minimum height=3em, minimum width=9em]
\tikzstyle{blockS} = [draw, fill=blue!20, rectangle, 
    minimum height=3em, minimum width=4em]    
\tikzstyle{input} = [coordinate]
\tikzstyle{output} = [coordinate]
\usetikzlibrary{matrix}

\usetikzlibrary{positioning}
\usetikzlibrary{math}

\usepackage{hyperref}
\usepackage{xcolor}
\hypersetup{
    colorlinks,
    linkcolor={blue!100!black},
    citecolor={blue!50!black},
    urlcolor={blue!80!black}
}

\newcommand{\bc}{\begin{center}}
\newcommand{\ec}{\end{center}}
\newcommand{\benum}{\begin{enumerate}}
\newcommand{\eenum}{\end{enumerate}}
\newcommand{\nn}{\nonumber}
\newcommand{\matl}{\left[ \begin{array}}
\newcommand{\matr}{\end{array} \right]}
\newcommand{\matls}{\left[ \begin{smallmatrix}}
\newcommand{\matrs}{\end{smallmatrix} \right]}
\newcommand{\isdef}{\stackrel{\triangle}{=}}

\newcommand{\rmN}{{\rm N}}
\newcommand{\rmO}{{\rm O}}

\newcommand{\rmT}{{\rm T}}

\newcommand{\rmc}{{\rm c}}

\newcommand{\rmf}{{\rm f}}

\newcommand{\rmp}{{\rm p}}

\newcommand{\BBR}{{\mathbb R}}

\newcommand{\SO}{{\mathcal O}}

\newcommand{\SSS}{{\mathcal S}}

\newcommand{\shiftq}{{\textbf{\textrm{q}}}}

\newcommand{\scM}{{\mathcal M}}

\newcommand{\EndProof}{$\hfill\mbox{\large$\square$}$}

\newcommand{\EndExample}{$\hfill\mbox{\Large$\diamond$}$}

\usepackage[style=numeric-comp ,sorting=none,maxbibnames=99,giveninits]{biblatex}
\usepackage{fancyhdr}
\usepackage{xcolor}
\newcommand{\redtext}[1]{\textcolor{red}{#1}}
\renewcommand{\redtext}[1]{\textcolor{black}{#1}}
\begin{document}

\author{Ankit Goel\thanks{Research Fellow, Aerospace Engineering Department. \texttt{ankgoel@umich.edu}} 
and Dennis S. Bernstein\thanks{Professor, Aerospace Engineering Department. \texttt{dsbaero@umich.edu}}}
\institute{University of Michigan, Ann Arbor}
\title{Retrospective Cost Parameter Estimation  with Application to Space Weather Modeling}
\maketitle

\begin{abstract}
    \redtext{
    This chapter reviews standard parameter-estimation techniques and presents a novel gradient-, ensemble-, adjoint-free data driven parameter estimation technique in the DDDAS framework.
    This technique, called retrospective cost parameter estimation (RCPE), is motivated by large-scale complex estimation models characterized by high-dimensional nonlinear dynamics, nonlinear parameterizations, and representational models.  
    %
    %
    RCPE is illustrated by estimating unknown parameters in three examples. 
    In the first example, salient features of RCPE are investigated by considering parameter estimation problem in a low-order nonlinear system. 
    In the second example, RCPE is used to estimate the convective coefficient and the viscosity in the generalized Burgers equation by using a scalar measurement.
    In the final example, RCPE is used to estimate thermal conductivity coefficients that relate temporal temperature variation with vertical gradient of the temperature in the atmosphere. 
    }
\end{abstract}

\section{Introduction}

The upper atmosphere, also called the thermosphere, is a complex system driven by solar activity on its outer edge and the oceanic and atmospheric activities on the inner edge.  
This region is characterized by high-temperature gases due to the absorption of high-energy solar radiation between about 80 km and about 600 km above sea level. 
Photo-ionization and photo-dissociation due to ultraviolet radiation create ions, and thus a large part of the thermosphere is ionized.   
Furthermore, since the atmospheric gases sort themselves in the thermosphere according to their molecular mass, the atmospheric density varies with  altitude.

Understanding the thermospheric dynamics is crucial for various research and industrial applications such as trajectory tracking of satellites and improving the accuracy of GPS signals.
Better density estimates in the upper atmosphere improve satellite tracking accuracy, whereas better knowledge of total electron content and its gradient improves  GPS  accuracy.
Furthermore,  improved understanding of the thermosphere provides insight into the climate of the lower atmosphere and the magnetosphere activities.
Global models of the thermosphere and data collected by various satellites and ground stations facilitate prediction of  the complex processes in the thermosphere.

Advances in high-performance computing and parallel computing have enabled numerical simulation of complex, large-scale, and high-dimensional systems like the upper atmosphere. 
One such model is the Global Ionosphere Thermosphere Model (GITM) \cite{ridley2006global}.
GITM is a computational code that models the thermosphere and the ionosphere of the Earth as well as that of various planets and moons by solving coupled continuity, momentum, and energy equations. 
By propagating the governing equations, GITM computes neutral, ion, and electron temperatures, neutral-wind and plasma velocities, and mass and number densities of neutrals, ions, and electrons. 
GITM uses empirical models, closure models, and sub-grid models to account for the physics of the atmospheric phenomenology affording high-resolution estimates of states.
In a typical simulation, GITM has more than 10 million states.
This chapter considers the problem of estimating parameters in the high-dimensional models like GITM to support advances in Dynamic Data Driven Applications Systems (DDDAS).


The chapter is organized as follows. %
In Section \ref{sec:ProblemFormulation}, the parameter estimation problem with appropriate assumptions is posed. 
Section \ref{sec:traditionalMethods} describes methods traditionally applied to the parameter estimation problem in dynamical systems.
Section \ref{sec:RCPE} introduces a data-driven parameter estimation technique called the \textit{retrospective cost parameter estimation} (RCPE) algorithm. 
In Section \ref{sec:NumExmp_lu3_ly1}, salient features of RCPE are investigated by considering parameter estimation problem in a low-order nonlinear system.
In Section \ref{sec:Burgers}, RCPE is used to estimate the convective coefficient and the viscosity in the generalized Burgers equation by using a scalar measurement. 
In Section \ref{sec:GITM}, RCPE is used to estimate thermal conductivity coefficients that relate temporal temperature variation with vertical gradient of the temperature in the atmosphere using simulated density measurements.  
%

\section{Parameter-Estimation Problem}
\label{sec:ProblemFormulation}

Consider the discrete-time system  
\begin{align}
    x_{k+1}
        &=
            f ( x_k, u_k,  \mu ),
    \label{eq:NL_state}
    \\
    y_k
        &=
            g ( x_k, u_k, \mu ),
    \label{eq:NL_output}
\end{align}
where $x_k \in \mathbb{R}^{l_x}$ is the state,
$u_k \in \mathbb{R}^{l_{u}}$ is the measured input,
$y_k \in \mathbb{R}^{l_{y}}$ is the measured output, 
$\mu = [\mu_1\  \cdots\  \mu_{l_\mu}]^\rmT \in  \mathbb{R}^{l_\mu}$ is the {\it true parameter}, which is unknown.
The system \eqref{eq:NL_state}, \eqref{eq:NL_output} is viewed as the \textit{truth model} of a physical system. 
Based on \eqref{eq:NL_state}, \eqref{eq:NL_output}, the {\it estimation model} is constructed as
\begin{align}
    \hat x_{k+1}
        &=
            f ( \hat x_k, u_k,  \hat \mu ),
    \label{eq:NL_Obs_state}        
    \\
    \hat y_k
        &=
            g ( \hat x_k,  u_k, \hat{\mu} ),
    \label{eq:NL_Obs_Output}        
\end{align}
where $\hat x_k$ is the computed state, 
$\hat y_k$ is the computed output of \eqref{eq:NL_Obs_state}, \eqref{eq:NL_Obs_Output}, and $ \hat{\mu}_k$ is the \textit{parameter estimate}. 
It is assumed that $f$ and $g$ are known, and thus they can be used to construct \eqref{eq:NL_Obs_state}, \eqref{eq:NL_Obs_Output}.
%
%
Since $\mu$ is unknown, it is replaced by $\hat\mu$ in \eqref{eq:NL_Obs_state}, \eqref{eq:NL_Obs_Output}.
The objective is to estimate $ \mu$ based on the {\it output error} $z_k\in \mathbb{R}^{l_y}$ defined by
\begin{align}
    z_k
        \isdef
            \hat y_k - y_k.\label{zref}
\end{align}
The ability to estimate $\mu$ is based on the assumption that \eqref{eq:NL_state}, \eqref{eq:NL_output} is structurally identifiable \cite{bellman1970structural,grewal1976identifiabilityb,stanhope2014identifiability} and the data are sufficiently persistent \cite{Mareels1987,willems2005note}. 
\section{Traditional Parameter Estimation Methods}
\label{sec:traditionalMethods}
Empirical models, closure models, and subgrid models are often of the form
\begin{align}
    \frac{\partial q}{\partial t} 
        =
             \lambda \frac{\partial q}{\partial x}   + Q, 
    \label{eq:emp_model}
\end{align}
where $q$ is the variable of interest, such as temperature or pressure,
$Q$ is a source term, and $\lambda$ is a coefficient that is fit using data. 
When the terms in \eqref{eq:emp_model} cannot be measured, due to lack of instrumentation or are \redtext{a hypothetical model to account for unmodeled physics} and thus cannot be measured, regression techniques cannot be used to interpolate the missing variables.

One approach to parameter fitting is to simulate the model \eqref{eq:NL_Obs_state}, \eqref{eq:NL_Obs_Output} multiple times with various choices of $\hat \mu.$.
By comparing the output of each simulation with the measurements,
the parameter value $\hat \mu$ that yields the output closest to the measurements is then accepted as the parameter estimate. 
This approach is impractical as the model size increases.
Furthermore, lack of knowledge of the initial conditions and disturbances may lead to an incorrect value of the unknown parameter that yields an output closest to the measurement, thus making this approach unreliable.

A more systematic approach to parameter fitting is to minimize the cost function
\begin{align}
    J(\hat \mu)
        \isdef 
            \sum_{i=1}^N 
            (y_i - \hat y_i )^\rmT
            W_i
            (y_i - \hat y_i ),
    \label{eq:cost}
\end{align}
where
$W_i \in \BBR^{l_y \times l_y}$ is a positive-definite weighting matrix. 
%
The complexity of minimizing \eqref{eq:cost} depends on the model \eqref{eq:NL_Obs_state}, \eqref{eq:NL_Obs_Output} of the system. 
\textit{Regression techniques} can be used in the case where the output $y_k$ is linearly parameterized by $\mu$, that is, 
\begin{align}
    y_k
        = 
            \phi_k \mu,
    \label{eq:lin_reg_form}
\end{align}
and the regressor $\phi_k \in \BBR^{l_y \times l_\mu}$ is measured.
With the measured regressor, a closed-form analytical expression for the minimizer of \eqref{eq:cost} exists and  can be recursively computed using the \textit{recursive least square} (RLS) algorithm to estimate $\mu$ \cite{AseemRLS, AppliedSystemIdentification_book, ljung:83,ljung_sysID_book}.
Additionally, if the regressor sequence $\{\phi_k\}_{k\ge0}$ is persistently exciting, then the RLS estimate converges to $\mu$ geometrically at a rate that can be made arbitrarily fast using appropriate forgetting \cite{goel2020_VDF}.  
Since the RLS estimate is sequentially updated with the available measurements, this method can be used online to estimate the unknown parameter $\mu.$

In the case where the model \eqref{eq:NL_Obs_state}, \eqref{eq:NL_Obs_Output} cannot be reformulated in the linearly parameterized form \eqref{eq:lin_reg_form}, an iterative optimization technique such as  gradient descent can be used to estimate the parameter \cite{boyd2004convex}. 
Using the \textit{gradient descent method}, the parameter estimate at the $j$th iteration is given by
\begin{align}
    \hat \mu(j) 
        = 
            \hat \mu(j-1) + \gamma 
            \left. \frac{\partial J}{\partial \mu} \right|_{\mu = \hat \mu(j-1)}
            .
    \label{eq:grad_desc_algo}
\end{align}
A first-order approximation of the gradient at $\mu=\hat \mu$ to be used in \eqref{eq:grad_desc_algo} can be computed by
\begin{align}
    \left. \frac{\partial J}{\partial \mu_i} \right|_{\mu = \hat \mu} 
        \approx
            \frac{ J(\hat \mu+\delta e_i)-J(\hat \mu-\delta e_i)}{2\delta},
    \label{eq:grad_approx}
\end{align}
where $\delta>0,$  $i \in \{ 1,2, \ldots, l_\mu \},$ and $e_i$ is the $i$th column of the identity matrix $I_{l_\mu}.$
Note that the estimation model \eqref{eq:NL_Obs_state}, \eqref{eq:NL_Obs_Output} needs to be run $2 l_\mu$ times to collect $N$ output vectors $\hat y_i$ required to compute the gradient given by \eqref{eq:grad_approx} at each iteration. %
Thus, each iteration of the gradient descent method \eqref{eq:grad_desc_algo} requires $2 l_\mu N$ executions of the functions $f$ and $g.$
Furthermore, the accuracy of this method may depend on the initial conditions of the estimation model \eqref{eq:NL_Obs_state}, \eqref{eq:NL_Obs_Output}. 
If the system is asymptotically stable, then the effect of the initial condition can be reduced by increasing $N$, that is, considering a large data set. 
On the other hand, if the effect of the initial condition persists, then the estimate may be biased even as $N \to \infty.$
%
For continuous-time models, the method of adjoints can be used to compute the gradient used in \eqref{eq:grad_desc_algo} \cite{kouba_adjoint_2020,jaakkola2000bayesian,raffard2008adjoint,eknes1997parameter}. 

Note that the gradient is computed by processing all of the data at once. 
Since the gradient computation requires the entire output sequence, the model \eqref{eq:NL_Obs_state}, \eqref{eq:NL_Obs_Output} is simulated for $N$ steps at each iteration, and therefore this method  cannot be used online to estimate $\mu.$

Alternatively, state estimation techniques can be used to estimate unknown parameters by modeling them as constant states. 
Since these parameters multiply dynamic states, the corresponding state estimation problem is nonlinear. 
%
%
First, the augmented state is defined as
\begin{align}
    X_k
        \isdef
            \matl{c}
                x_k \\
                \mu
            \matr.
\end{align}
Define $\pi_1 \isdef [I_{l_x} \ 0_{l_x \times l_\mu}]$ and
$\pi_2 \isdef [0_{l_\mu \times l_x} \  I_{l_\mu}]$ so that 
$x_k = \pi_1 X_k$ and $\mu_k = \pi_2 X_k$.
The augmented dynamics is
\begin{align}
    X_{k+1} 
        =
            F(X_k,u_k), 
    \label{eq:AugState}
        \\
    y_k
        =
            G(X_k,u_k),
    \label{eq:AugOutput}
\end{align}
where 
\begin{align}
    F(X_k,u_k)
        &\isdef 
            \matl{c}
                f( \pi_1 X_k, u_k, \pi_2 X_k) \\
                \pi_2 \mu_k
            \matr,
        \\
    G(X_k,u_k)
        &\isdef 
                g(\pi_1 X_k, u_k, \pi_2 X_k),
\end{align}
%
%
%
Nonlinear extensions of the classical Kalman filter \cite{simon2006optimal} such as the extended Kalman filter (EKF), the ensemble Kalman filter (EnKF), and the unscented Kalman filter (UKF) can be used to estimate the augmented state \cite{Ljung1979,plett2004extended,van2001square,wan2000unscented,evensen2009ensemble,moradkhani2005dual,madankan_2013}.
The parameter estimate $\hat \mu_k$ is then given by
\begin{align}
    \hat \mu_k
        =
            \pi_2 \hat X_k,
    \label{eq:muhat_k_St_est}
\end{align}
where $\hat X_k \isdef [\hat x_k^\rmT \ \hat \mu_k^\rmT]^\rmT $ is the augmented state estimate.

The extended Kalman filter requires the computation of the gradient of $F$ and $G$ with respect to the augmented state.
However, the gradient computation, both analytical and numerical, becomes prohibitively computationally expensive as the model size increases.

Ensemble-based methods such as EnKF, UKF, and the particle filter, while obviating the need for gradient computation, require making copies of the state, adding appropriate perturbations to populate the ensemble, and finally propagating each ensemble member using the estimation model based on \eqref{eq:AugState}, \eqref{eq:AugOutput}.  
To efficiently utilize the computational resources, typical implementations of such estimation models often do not save the state at each step but overwrite the state.
Furthermore, the state update might not be a batch operation; the state's entries might be updated sequentially in subiterations as is common in implicit numerical methods \cite{hoffman2018numerical}. 
In addition to the computational cost of multiple executions of the functions $f$ and $g$ at each step, the states need to be copied and indexed appropriately, and thus these techniques require considerable programming effort. 
Consequently, the implementation of ensemble-based methods is specific to the application, and ensemble-based methods are thus not modular. 
%

%

%
%

\section{Retrospective Cost Parameter Estimation}
\label{sec:RCPE}

The difficulties outlined in the previous section motivate the need for a modular, gradient-, ensemble-free, data-driven parameter estimation technique as described in this section. 
Based on the retrospective cost adaptive control \cite{Rahman2017}, this technique, called the \textit{retrospective cost parameter estimation} (RCPE) algorithm uses past data to estimate the unknown parameter.  \cite{goel2019gradient,goel2020estimation_TC,goel2018data,ankit_EDC_ACC2018,goel2016parameter}. 
RCPE uses the estimation model 
\begin{align}
    \hat x_{k+1} 
        &=
            f(\hat x_k,u_k,\hat \mu_k), 
    \label{eq:model_state}
    \\
    \hat y_k
        &=
            g(\hat x_k,u_k,\hat \mu_k),
    \label{eq:model_output}
\end{align}
where $\hat \mu_k$ is the parameter estimate to compute the output $\hat y_k.$
The error signal $z_k$ given by the difference between the output $y_k$ of the physical system and the output $\hat y_k$ of the estimation model is used to minimize a retrospective cost function whose minimizer is used to construct the parameter estimate $\hat \mu_k$ .

%
%

RCPE assumes that the unknown parameter satisfies 
$\mu = [\mu_1\  \cdots\  \mu_{l_\mu}]^\rmT \in \scM,$
%
where the set $\scM$ is assumed to be known and satisfy 
$\scM\subseteq[0,\infty)^{l_\mu},$ that is, $\scM$ is contained in the nonnegative orthant. 
If  $\scM$ does not satisfy this condition, then it may be possible to replace $\scM$ by $\scM'\isdef\overline{\mu}+\scM$ and $\mu$ by $\mu-\overline{\mu}$ in 
\eqref{eq:NL_state}, \eqref{eq:NL_output}, 
where $\overline{\mu}\in\BBR^{l_\mu}$ shifts $\scM$ such that $\scM'$ is contained in  the nonnegative orthant.
With this transformation, which can always be done if $\scM$ is bounded, it can be assumed that $\mu$ is an element of  the nonnegative orthant.

Like UKF, but unlike EKF, RCPE does not require a Jacobian of the dynamics in order to update the parameter estimates. 
However, unlike UKF, RCPE does not require an ensemble of models.
In particular, for parameter estimation, the UKF is based on an ensemble of $2n+1$ models, where $n=l_x+l_\mu,$ $l_x$ is number of dynamic states, and $l_\mu$ is the number of unknown parameters.
The total number of states that must be propagated at each iteration is thus $2n^2+n = O(n^2).$
%
%
In contrast to UKF, RCPE requires the propagation of only a single copy of the ``original'' system dynamics, so that the number of states that must be propagated at each iteration is simply $l_x.$
For both UKF and RCPE, this model need only be given as an executable simulation; explicit knowledge of the  equations and source code underlying the simulation is not required. 
However, the price paid for not requiring an explicit model or an ensemble of models is the need within RCPE to select a permutation matrix that correctly associates each parameter estimate with the corresponding unknown parameter. 
%
%

%

\subsection{Parameter Estimator}
\label{sec:parameter_estimator}

\begin{figure}[ht]
	\centering
	    \resizebox{\columnwidth}{!}{
 		\begin{tikzpicture}[auto, node distance=2cm,>=latex']

			\node at (0,0) [ultra thick, block, align=center, node distance=0 cm] (MainSystem) 
			{
			    $\begin{array}{c} 
			    x_{k+1}
                    =
                        f ( x_k, u_k,  \mu )
                \\
                \quad y_k
                    =
                        g ( x_k, u_k, \mu )
                \end{array}
                $
            };
			
			\node [ultra thick, block, align=center, below of = MainSystem, node distance = 2 cm](MainSystemModel) 
            {
                $\begin{array}{c} 
			    \hat x_{k+1}
                    =
                        f ( \hat x_k, u_k,  \hat \mu_k )
                \\
                \quad \hat y_k
                    =
                        g ( \hat x_k, u_k, \hat \mu_k )
                \end{array}
                $
            };
			
			\node [ultra thick, block, align=center, below of = MainSystemModel, node distance = 2 cm](RCPE) 
			{$
			\begin{array}{c}
			    \Phi_k = \Phi_{k-1} + I_{l_\mu} \otimes z_{k-1}^\rmT
			    \\
			    \hat \mu_k = \SO_\rmp | \Phi_k \theta_k |
			\end{array}
			$};
			
			\node [ultra thick, sum, align=center, right of = MainSystemModel, node distance = 3 cm] (sum) {}; 
			\node at (RCPE)      [xshift=0.5 cm, yshift=-1 cm] (belowRCPE) {};
			\node at (RCPE)      [xshift=-0.5 cm, yshift=1.2 cm] (leftofRCPE) {};
			
			\draw [ultra thick,->] (-3,0) node [xshift=0.5 cm, yshift=8 pt] {$u$}  
			    |- (MainSystem.west) ;
			\draw [ultra thick,->] (-2.5,0)     |- (MainSystemModel.175) ;			
			
			\draw [ultra thick,->] (MainSystem.east) node [xshift=.6cm, yshift=8pt] {$y$} 
			        -|  (sum.north) node [xshift=5 pt, yshift=0pt] {$-$} ;
			\draw [ultra thick,->] (MainSystemModel.east) node [xshift=.6cm, yshift=8pt] {$\hat{y}$} 
			        --  (sum.west) ;			
			\draw [ultra thick,->] (sum.south) |-  (RCPE.east) node [xshift=.5cm, yshift=8pt] {$z$} ;
			\draw [ultra thick,->] (sum.south) |-  (belowRCPE.center) -- (leftofRCPE);
			\draw [ultra thick,->] (RCPE.west) node [xshift=-0.35cm, yshift=8pt] {$\hat \mu$}  -- (-2.5,-4) |-  (MainSystemModel.185);

			\node [ultra thick, block, align=center, below of = MainSystemModel, node distance = 2 cm](RCPE) 
			{$
			\begin{array}{c}
			    \Phi_k = \Phi_{k-1} + I_{l_\mu} \otimes z_{k-1}^\rmT
			    \\
			    \hat \mu_k = \SO_\rmp | \Phi_k \theta_k |
			\end{array}
			$};

		\end{tikzpicture}
		}
		\caption
		{
			Retrospective cost parameter estimation. %
			The output error $z$ is used to optimize the parameter estimator, whose output is the parameter estimate $\hat \mu.$
		}
	\label{fig:RCPE_formulation}
\end{figure}
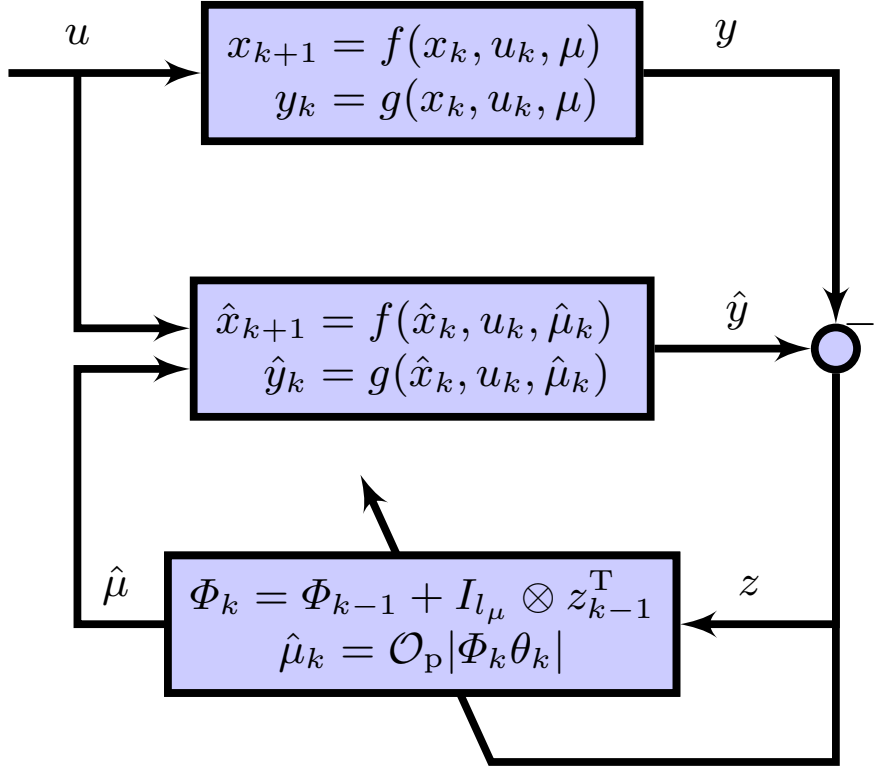

The {\it parameter estimator} consists of an adaptive integrator and an output nonlinearity.  
In particular, the {\it parameter pre-estimate} $\nu$ is given by
\begin{align}
	{\nu}_k 
	    =
		    R_k \phi_k,
	\label{controller}
\end{align}
where the \textit{integrator state} $\phi_k\in\BBR^{l_y}$ is updated by
\begin{align}
    \phi_k 
        =
            \phi_{k-1} + z_{k-1}.
    \label{eq:integratorDef}
\end{align}
The \textit{adaptive integrator gain} $R_k\in \mathbb{R}^{l_\mu \times l_{y}}$ is updated by RCPE as described later in this section.
Since $\nu_k$ is not necessarily an element of the nonnegative orthant, an output nonlinearity is used to transform $\nu_k$.
In particular, the { parameter estimate} $\hat\mu_k$ is given by   
\begin{align}
    \hat \mu_k
        = 
            \SO_\rmp | \nu_k |,
    \label{eq:AAVT_map}
\end{align}
where the absolute value is applied componentwise.
The matrix $\SO_\rmp$ is explained below. 
The parameter estimator, which consists of \eqref{controller},  \eqref{eq:integratorDef}, \eqref{eq:AAVT_map}, is represented in Figure  \ref{fig:RCPE_formulation}.
Since $z_k \to 0$ as $k\to\infty$  is a necessary condition for $\phi$ to converge, the integrator \eqref{eq:integratorDef} allows $z$ to converge to zero while $\phi$ converges to a finite value.
Consequently, the parameter pre-estimate $\nu$ given by \eqref{controller} can converge to a nonzero value, which, in turn, allows the parameter estimate $\hat\mu,$ given by \eqref{eq:AAVT_map}, to converge to $\mu$. 

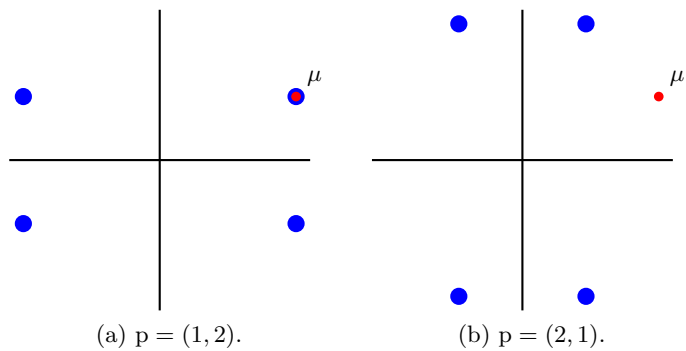
\begin{figure}[ht]
	\centering
	\subfloat[$\rmp = (1,2).$]
	    {
	        \begin{tikzpicture}[auto, node distance=2cm,>=latex']
			\tikzmath
                {
                    \Rw     = 2;
                    \ang    = 25;
                } 
            
            \draw [thick] (-\Rw,0) -- (\Rw,0);
            \draw [thick] (0,-\Rw) -- (0,\Rw);

		    \draw [thick, fill=black, blue] 
			    ({\Rw*cos(\ang)},{\Rw*sin(\ang)}) circle [radius=.1];
		    \draw [thick, fill=black, blue] 
			    ({\Rw*cos(\ang)},{-\Rw*sin(\ang)}) circle [radius=.1];
			\draw [thick, fill=black, blue] 
			    ({-\Rw*cos(\ang)},{-\Rw*sin(\ang)}) circle [radius=.1];
            \draw [thick, fill=black, blue] 
			    ({-\Rw*cos(\ang)},{\Rw*sin(\ang)}) circle [radius=.1];	

			\draw [thick, red, fill=red] 
			    ({\Rw*cos(\ang)},{\Rw*sin(\ang)}) 
			    node[xshift=7,yshift=7, black] {$\mu$}
			    circle [radius=.05];
		\end{tikzpicture}
		\label{fig:SetS_lu2_p12}
        }
        \quad 
        \subfloat[$\rmp=(2,1).$]
	    {
	        \begin{tikzpicture}[auto, node distance=2cm,>=latex']
			\tikzmath
                {
                    \Rw     = 2;
                    \ang    = 25;
                } 
            
            \draw [thick] (-\Rw,0) -- (\Rw,0);
            \draw [thick] (0,-\Rw) -- (0,\Rw);
			\draw [thick, red, fill=red] 
			    ({\Rw*cos(\ang)},{\Rw*sin(\ang)}) 
			    node[xshift=7,yshift=7, black] {$\mu$}
			    circle [radius=.05];
			\draw [thick, fill=black, blue] 
			    ({\Rw*sin(\ang)},{\Rw*cos(\ang)}) circle [radius=.1];
		    \draw [thick, fill=black, blue] 
			    ({\Rw*sin(\ang)},{-\Rw*cos(\ang)}) circle [radius=.1];
			\draw [thick, fill=black, blue] 
			    ({-\Rw*sin(\ang)},{-\Rw*cos(\ang)}) circle [radius=.1];
            \draw [thick, fill=black, blue] 
			    ({-\Rw*sin(\ang)},{\Rw*cos(\ang)}) circle [radius=.1];	
		\end{tikzpicture}
		\label{fig:SetS_lu2_p21}
        }
		\caption
		{
			The set $\SSS_{\SO_\rmp}$ for $l_\mu = 2$ consists of the blue dots; 
			$\mu$ is shown in red. 
		}
	\label{fig:SetS_lu2}
\end{figure}

Let the $l_\mu$-tuple $\rmp = (i_1,\ldots,i_{l_\mu})$ denote a permutation of $(1, \ldots, l_\mu )$.
Then the matrix $\SO_\rmp \in\BBR^{l_\mu\times l_\mu}$ maps $(1, \ldots, l_\mu )$ to $(i_1,\ldots,i_{l_\mu}).$
Since $\SO_\rmp$ is a permutation matrix, each of its rows and columns contains exactly one ``1,'' and the remaining entries are all zero.
Specifically,  row $j$ of $\SO_\rmp$ is row $i_j$  of the identity matrix $I_{l_\mu}$.
Now, define the set 
\begin{align}
    \SSS_{\SO_\rmp} \isdef  \{s\in\BBR^{l_\mu}\colon \SO_\rmp|s| = \mu \},
    \label{eq:S_Op_def}
\end{align} 
whose elements are the vectors that are mapped to $\mu$ by the componentwise absolute value and the permutation $O_\rmp$.
For illustration, Figure \ref{fig:SetS_lu2_p12} shows the elements of $\SSS_{\SO_{12}}$, and Figure \ref{fig:SetS_lu2_p21} shows the elements of $\SSS_{\SO_{21}}$.

In the following development, parameter pre-estimate \eqref{controller} is rewritten as
\begin{align}
	{\nu}_k = \Phi_k \theta_k,
	\label{eq:nu_regressor_form}
\end{align}
where the regressor matrix $\Phi_k$ is defined by
\begin{align}
	\Phi_k \isdef
	    I_{l_\mu}
	    \otimes 
		\phi^ {\rm T} _k 
		\in \mathbb{R}^{l_\mu \times l_{\theta}}, 
	\label{eq:Phi_def}
\end{align}
\redtext{$I_{l_\mu}$ is the $l_\mu \times l_\mu$ identity matrix}, 
and the estimator coefficient $\theta_k$ is defined by
\begin{align}
    \theta_k 
        &\isdef
            \textrm{vec}\ 
            R_k
        \in \mathbb{R}^{l_{\theta}},    %
\end{align}
where
$l_\theta \isdef l_\mu l_y$,
``$\otimes$'' is the Kronecker product, and ``vec'' is the column-stacking operator. %
Note that $\theta_k$ is an alternative representation of the adaptive integrator gain $R_k$.
%

\subsection{Retrospective Cost Optimization}
The {\it retrospective error variable} is defined by
\begin{align}
    {\hat z_k(\hat \theta)}
        \isdef
            z_k + G_\rmf{(\rm{\bf q})} 
            [ 
                \Phi_k \hat{\theta} - \nu _k 
            ],
    \label{eq:z_hat_def}
\end{align}
where $\shiftq$ is the forward-shift operator and 
$\hat\theta\in\BBR^{l_\theta}$ is determined by optimization to obtain the updated estimator coefficient $\theta_{k+1}.$
The filter $G_\rmf$ has the form
\begin{align}
    G_\rmf(\shiftq) 
        =
     \sum_{i=1}^{n_\rmf}\frac{1} {\shiftq^i}N_i, 
\end{align}
where $N_1,\ldots,N_{n_\rmf} \in \BBR^{l_y \times l_\mu}$ are the \textit{filter coefficients}, \redtext{and thus,} $G_\rmf$ is an $l_y \times l_\mu$ finite impulse response filter.
\redtext{Note that, for all $i \ge0,$ $\shiftq^i \nu_k = \nu_{k+i}$.}
%
%
The retrospective error variable \eqref{eq:z_hat_def} can thus be rewritten as
\begin{align}
    {\hat z_k(\hat \theta)}
        =
            z_k + N \overline \Phi_k \hat{\theta} - N \overline V _k,
\end{align}
where
\begin{gather}
    N
        \isdef 
            [
                N_1 \ \cdots \ N_{n_\rmf}
            ] 
            \in \BBR^{l_y \times n_\rmf l_\mu},
            \label{eq:N_def}
            \\
    \overline \Phi_k
        \isdef 
            \matl{c}
                \Phi_{k-1} \\
                \vdots \\
                \Phi_{k-n_\rmf}
            \matr
            \in \BBR^{l_\mu n_\rmf \times l_\theta},
            \quad   
    \overline V_k
        \isdef 
            \matl{c}
                \nu_{k-1} \\
                \vdots \\
                \nu_{k-n_\rmf}
            \matr
            \in \BBR^{l_\mu n_\rmf }.
            \label{eq:Phibar_Vbar_def}
\end{gather}
%

The \textit{retrospective cost function} is defined by
\begin{align}
    J_k(\hat{\theta}) 
	    \isdef& 
	        \sum_{i=1}^{k} 
	            \lambda^{k-i} {\hat z_i(\hat \theta)}^{\rm T} {\hat z_i(\hat \theta)} +
	            \lambda^k \hat \theta ^{\rm T} R_{\theta}  \hat \theta,
\label{Jg}
\end{align}
where $R_{\theta} \in \BBR^{l_\theta \times l_\theta}$ is  positive definite and $\lambda \in (0,1]$ is the forgetting factor.
\redtext{
For all $k\ge0,$ the estimator coefficient $\theta_{k+1}$ is set equal to the minimizer of \eqref{Jg},
that is,
\begin{align}
    \theta_{k+1}
        =
            \underset{ \hat\theta \in \BBR^\theta  }{\operatorname{argmin}} \
            J_k({\hat\theta}).
    \label{eq:theta_minimizer_def}
\end{align}
}
The following result uses recursive least squares (RLS) to minimize \eqref{Jg}.

\begin{proposition}
    Let $P_0=R_\theta^{-1}$ and $\theta_0=0$.
    Then, for all $k \ge 1$, the retrospective cost function \eqref{Jg} has a unique global minimizer $\theta_{k+1}$, which is given by 
    \begin{align}
        P_{k+1}
            &=
                \lambda^{-1} [P_k - 
                P_k \overline \Phi_k^\rmT N^\rmT \Gamma _k^{-1}
                N \overline \Phi_k P_k],
        \label{eq:PUpdate}   
        \\
        \theta_{k+1}
            &=
                \theta_k -
                P_{k+1} \overline \Phi_k^\rmT N^\rmT 
                [
                    N \overline \Phi_k \theta_k + z_k - N \overline V_k
                ],
        \label{eq:thetaUpdate}
     \end{align}        
     where   
        \begin{align}
        \Gamma_k
            \isdef 
                \lambda I_{l_y} + N \overline \Phi_k P_k \overline \Phi_k^\rmT N^\rmT
                .
        \label{eq:GammaUpdate}
    \end{align}
\end{proposition}

Finally, the parameter estimate at step $k+1$ is given by
\begin{align}
    \hat \mu_{k+1}
        =
            \SO_\rmp |\nu_{k+1}|
        =
            \SO_\rmp |\Phi_{k+1} \theta_{k+1}|.
\end{align}
Since $\theta_0 = 0$, it follows that $\nu_0 = 0$ and thus $\hat \mu_0 = 0$.
\redtext{
Note that, since $z_k$ is not available until step $k,$ $\theta_{k+1}$ is computed between steps $k$ and $k+1$.
Consequently, $\theta_{k+1}$ is available at step $k+1$ and can be used in the estimation model \eqref{eq:NL_Obs_state}, \eqref{eq:NL_Obs_Output} at step $k+1$.
}

\subsection{The filter $G_\rmf$}
This section analyzes the role of the filter $G_\rmf$ in the update of the parameter pre-estimate $\nu$.
In particular, it is shown that the filter coefficients determine the subspace of $\BBR^{l_\mu}$ that contains $\nu$.

To analyze the role of $G_\rmf,$  the cost function \eqref{Jg} is rewritten as
\begin{align}
    J_k(\hat{\theta})
        &=
	        \hat{\theta}^\rmT A_{\theta  k} \hat \theta
	        +
	        2 b_{\theta  k}^\rmT \hat{\theta}
	        +
	        c_{\theta  k}, 
    \label{eq:RCAC_cost}
\end{align}
where
\begin{align}
    A_{\theta  k}
        &\isdef
            \sum_{i=1}^{k} 
                \lambda^{k-i}
                \overline \Phi_i^\rmT N^\rmT 
                N \overline \Phi_i + 
                \lambda^k
                R_\theta,
        \\
    b_{\theta  k}
        &\isdef
            \sum_{i=1}^{k} 
                \lambda^{k-i}
                \overline \Phi_i^\rmT N ^\rmT 
                (
                    z _i - N \overline V_i
                ),
    \label{eq:cumm_cost_b_theta} 
        \\
    c_{\theta  k}
        &\isdef
            \sum_{i=1}^{k} 
                \lambda^{k-i}
                ( z _i - N \overline V_i)^\rmT
                ( z _i - N \overline V_i ).
\end{align}
At step $k$, the batch least squares minimizer $\theta_{k+1}$ of \eqref{Jg} is given by
\begin{align}
    \theta_{k+1}
        =
            - A_{\theta  k}^{-1} b_{\theta  k},
    \label{eq:theta_update}
\end{align}
which is equal to 
$\theta_{k+1}$ given by \eqref{eq:thetaUpdate}.

The following result shows that the parameter pre-estimate $\nu_k$, and thus the estimate $\hat \mu_k$, is constrained to lie in a subspace determined by the coefficients of $G_\rmf.$

\begin{lemma}
    Let $\beta>0$, $R_\theta = \beta I_{l_\theta}$, 
    $\nu_k$ be given by \eqref{eq:nu_regressor_form}, 
    $\Phi_k$ be given by \eqref{eq:Phi_def}, 
    $N, \overline{\Phi}_k, \overline{V}_k$ be given by \eqref{eq:N_def}, 
    \eqref{eq:Phibar_Vbar_def}, 
    and 
    $\theta_{k+1}$ be given by \eqref{eq:theta_update}.
    Then, for all $k \ge 1$, 
    \begin{align}
        \nu_{k+1}
            &=
                -\frac{1}{\beta}
                [N_1^\rmT\ \cdots\ N_{n_\rmf}^\rmT]
                \sum_{i=1}^{k}
                    \lambda^{-i}
                    \Psi_{k,i} 
                    [
                        z _i + N \overline \Phi _i \theta_{k+1} - N \overline V_i
                    ]
                \nn \\
                &
                    \in 
                \mathcal{R} 
                ([N_1^\rmT\ \cdots\ N_{n_\rmf}^\rmT]
                ),
    \label{eq:uinRangeN}
    \end{align}
    where
    \begin{align}
        \Psi_{k,i}
            &\isdef
                \matl{c}
                    \phi_{k+1}^\rmT \phi_{i-1}  \otimes I_{l_y} \\
                    \vdots \\
                    \phi_{k+1}^\rmT \phi_{i-{n_\rmf}}  \otimes I_{l_y} \\
                \matr .
    \end{align}
    \label{th:uinRangeN}
\end{lemma}

\begin{proof}
    Note that 
    \begin{align}
        \Phi_{k+1} A_{\theta  k} \theta_{k+1} 
            &=
                \sum_{i=1}^{k} 
                    \lambda^{k-i}
                    \Phi_{k+1}
                    \left(
                        \overline \Phi_i^\rmT N^\rmT   N \overline \Phi_i 
                    \right) 
                    \theta_{k+1} +
                    \lambda^{k}\beta \Phi_{k+1} \theta_{k+1}
                    \nn \\
            &=
                \sum_{i=1}^{k} 
                    \left(
                    \lambda^{k-i}
                    \sum_{j=1}^{n_\rmf}
                        ( I_{l_u} \otimes \phi_{k+1}^\rmT )
                        ( I_{l_u} \otimes \phi_{i-j} ) N_j^\rmT 
                    \right) 
                      N \overline \Phi _i \theta_{k+1} 
                     \nn \\ &\quad \quad
                     +
                    \lambda^{k} \beta \Phi_{k+1} \theta_{k+1}
                \nn \\
            &=
                \sum_{i=1}^{k}
                    \left(
                    \lambda^{k-i}
                    \sum_{j=1}^{n_\rmf}
                        N_j^\rmT  \phi_{k+1}^\rmT \phi_{i-j}  
                    \right)   N \overline \Phi _i \theta_{k+1} +
                    \lambda^{k} \beta \Phi_{k+1} \theta_{k+1}
                \nn \\
            &=
                [N_1^\rmT\ \cdots\ N_{n_\rmf}^\rmT]
                \sum_{i=1}^{k}
                    \lambda^{k-i}
                    \Psi_{k,i}
                      N \overline \Phi _i \theta_{k+1} +
                    \lambda^{k} \beta \Phi_{k+1} \theta_{k+1}
            \label{eq:u_in_N_LHS}
    \end{align}
    and
    \begin{align}
        \Phi_{k+1} b_{\theta  k}
            &=
                \Phi_{k+1}
                \sum_{i=1}^{k}
                    \lambda^{k-i}
                    \overline \Phi_i^\rmT N ^\rmT  
                    \left(
                        z _i - N  \overline V_i
                    \right) \nn \\
            &=
                [N_1^\rmT\ \cdots\ N_{n_\rmf}^\rmT]
                \sum_{i=1}^{k}
                    \lambda^{k-i}
                    \Psi_{k,i}  
                    \left(
                        z _i - N \overline V_i
                    \right).
            \label{eq:u_in_N_RHS}
    \end{align}
    Writing \eqref{eq:theta_update} as $A_{\theta  k} \theta_{k+1} = - b_{\theta  k},$ multiplying by $\Phi_{k+1}$, and using \eqref{eq:u_in_N_LHS} and \eqref{eq:u_in_N_RHS} yields \eqref{eq:uinRangeN}.
    \EndProof
\end{proof}

It follows from Lemma \ref{th:uinRangeN} that the parameter pre-estimate $\nu$ is constrained to lie in the subspace of $\BBR^{l_\mu}$ spanned by the coefficients of the filter used by RCPE. 
In view of Lemma \ref{th:uinRangeN}, in the case where $l_z=1$, $n_\rmf$ is set to be equal to $l_\mu,$
and each filter coefficient is chosen to be an element of 
$\{e_1, e_2, \ldots, e_{l_\mu} \}$, where $e_i$ is the $i^{\rm th}$ row of the identity matrix $I_{l_\mu}$.
For $l_z>1$, the filter coefficients must be selected such that 
$\mu \in \mathcal{R} ([N_1^\rmT\ \cdots\ N_{n_\rmf}^\rmT]) $.
\redtext{
Finally, note that $[N_1^\rmT\ \cdots\ N_{n_\rmf}^\rmT]$ is a $l_\mu \times n_\rmf l_y$ matrix, whereas
$N\isdef [N_1 \ \cdots\ N_{n_\rmf}]$ is an $l_y \times n_\rmf l_\mu$ matrix. }

%

\section{Low-order example}
\label{sec:NumExmp_lu3_ly1}
In this example, RCPE is used to estimate three unknown parameters in an affinely parameterized nonlinear system with one measurement.
    Consider the (3,3) type nonlinear system 
    \cite[p. 183]{Kulenovic2002}
    \begin{align}
        x_{k+1}
            &=
                \matl{c}
                    x_{2k}     \\
                    \dfrac
                        {\mu_1 + \mu_2 x_{2k} + \mu_3 x_{1k}}
                        {1 + 0.6 x_{2k} + 1.1 x_{1k}}
                \matr + 
                \matl{c}
                    0 \\
                    1
                \matr
                u_{k},
        \label{eq:NL_NLP_state_lu3}
        \\
        y_{k}
            &=
                x_{1k},
        \label{eq:NL_NLP_output_lu3}
    \end{align}
    where $\mu= [ \mu_1\ \  \mu_2\ \ \mu_3]^\rmT = [0.5\ \ 0.8\ \ 1.0]^\rmT $.
    Let the initial state be $x_0 = [ 10\ \ 10\ \ 10 ]^\rmT$ and the input $u_{k}$ be given by
    \begin{align}
        u_k
            =
                2 + \sum_{i=1}^{15} \sin \frac{2\pi i}{100} k.
        \label{eq:u_k_loworderexample}
    \end{align}
    \redtext{
    Note that the input $u_k$ given by \eqref{eq:u_k_loworderexample} is used along with the initial state $x_0$ to generate the measurements $y_k$.
    %
    Furthermore, the initial state $\hat x_0$ of the estimation model constructed using \eqref{eq:NL_NLP_state_lu3}, \eqref{eq:NL_NLP_output_lu3} is set to zero.
    The objective is to estimate $\mu_1, \mu_2,$ and $\mu_3$ using the scalar measurement $y_k$.  
    Hence, $l_\mu=3$ and $l_y=1.$
    }

\begin{figure}[ht]
	\centering
	\includegraphics[width= \columnwidth]
	{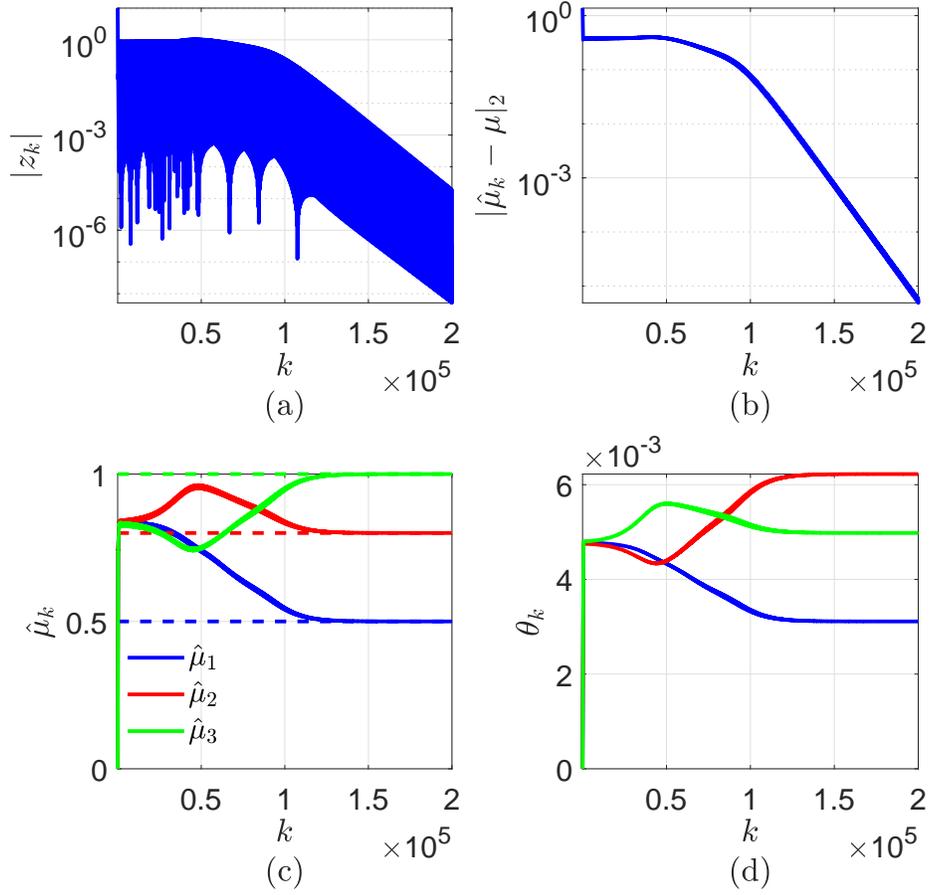}
    \caption 
    	{
    	    Estimation of three unknown parameters in the nonlinear system \eqref{eq:NL_NLP_state_lu3}, \eqref{eq:NL_NLP_output_lu3}.
    	    (a) shows the output error, 
    	    (b) shows the parameter error, 
    	    (c) shows the parameter estimate, and
    	    (d) shows the parameter estimator coefficients.
    	}
    \label{fig:DDDAS_Book_RCPE_NLS_NLP_lu3}
    \end{figure}

\begin{figure}[ht]
	\centering
	\includegraphics[width=1.0\columnwidth]{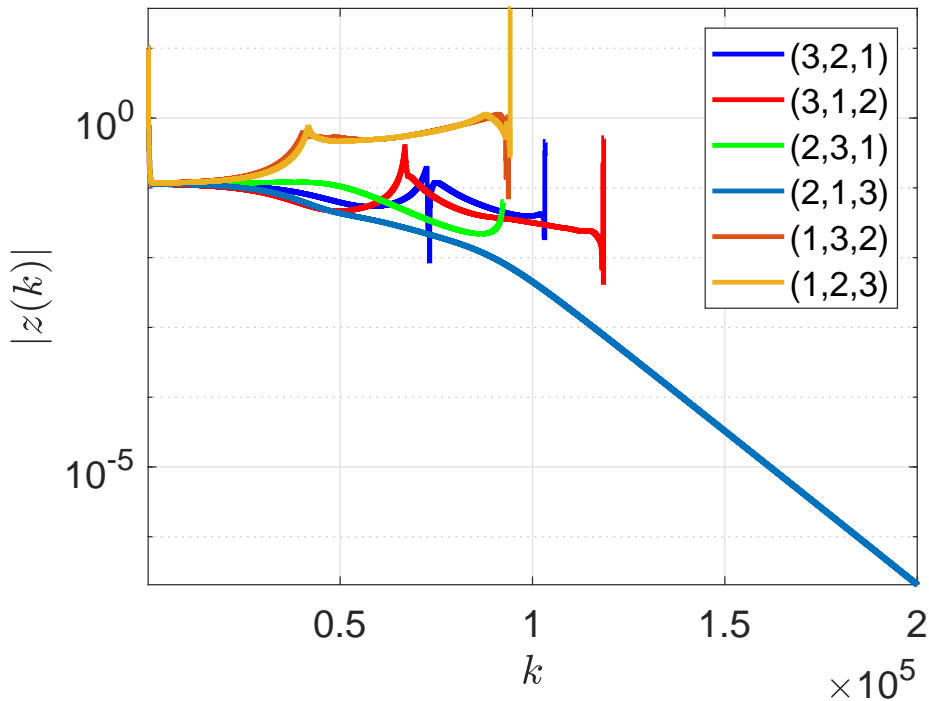}
    \caption 
        [Output error for all six permutations.]
    	{
    	    Estimation of three unknown parameters in the nonlinear system \eqref{eq:NL_NLP_state_lu3}, \eqref{eq:NL_NLP_output_lu3}.
    	    Output error for all six permutations. [For clarity, only a subset of the data is plotted.]
    	    For five of the six permutations, the parameter error diverges.  However, the remaining permutation (2,1,3) yields convergence to the true parameters.  
    	}
    \label{fig:RCMR_whitePaper_lu3_all_perms}
    \end{figure}

    To apply RCPE, let
    $N_1 = [1 \ 0 \ 0]$,
    $N_2 = [0 \ 1 \ 0]$,
    $N_3 = [0 \ 0 \ 1]$,
    $\lambda = 0.9999$, 
    $R_\theta = 10^6 I_3$, and 
    $\rmp = (2,1,3) $,
    and thus $\SSS_{\SO_\rmp} =  \{ [ \pm \mu_2 \  \pm \mspace{-4mu} \mu_1 \ \pm \mspace{-4mu} \mu_3]^\rmT  \}$.
    %
    %
    Figures \ref{fig:DDDAS_Book_RCPE_NLS_NLP_lu3} shows the output error, norm of the parameter error, the parameter estimates, and the estimator coefficients.
    Figure \ref{fig:RCMR_whitePaper_lu3_all_perms} shows the output error for all six permutations. 
    For clarity, a subset of the data is shown.
    Note that the output error diverges for all permutations except $(2,1,3)$.
    Therefore, diverging output error can be used to rule out the incorrect permutations. 

    Next, to investigate the relationship between the choice of the filter and the permutation matrix $\SO_\rmp$ that yields convergence, consider the filter
    \begin{align}
        G_{\rmf}(\shiftq)
            =
                \frac{\sigma_1 N_1}{\shiftq} +
                \frac{\sigma_2 N_2}{\shiftq} +
                \frac{\sigma_3 N_3}{\shiftq},
    \end{align}
    where, for $i=1,2,3,$ $N_i \in \{ e_1, e_2, e_3\}$ and $\sigma_i \in \{\pm 1 \}.$
    Note that $e_i$ is the $i$th row of the $3\times 3$ identity matrix. 
    Since there are six permutations of the set $\{ e_1, e_2, e_3\}, $
    there are six choice of the filter coefficients that satisfy Lemma \ref{th:uinRangeN}.
    %
    The choices of coefficients $N_1, N_2,$ and  $N_3$ 
    are given in  Table \ref{tab:filter_coefficients}.
    For each choice of the filter coefficients, exactly one permutation matrix yields convergence, which is shown in the fourth column of  Table \ref{tab:filter_coefficients}, whereas all other permutations lead to divergence of the output error $z_k$. 
    \redtext{Note that the correct permutation is obtained by checking all permutations in numerical simulations.}
    
    Furthermore, for each choice of the filter coefficients, there are eight possible combinations of $\sigma_1,$ $\sigma_2,$ and $\sigma_3,$
    %
    which are listed in  Table \ref{tab:filter_coefficients_sign}. 
    With the correct permutation matrix choice for each choice of the filter $G_\rmf(\shiftq)$, the coefficients $N_i$ are multiplied by $\sigma_i$  given in  Table \ref{tab:filter_coefficients_sign}.
    There are thus 48 possible filters. 

\begin{figure}[ht]
    	\centering
    	\includegraphics[width= 0.9\columnwidth, trim=2.2cm 0 1.8cm 0, clip]
    	{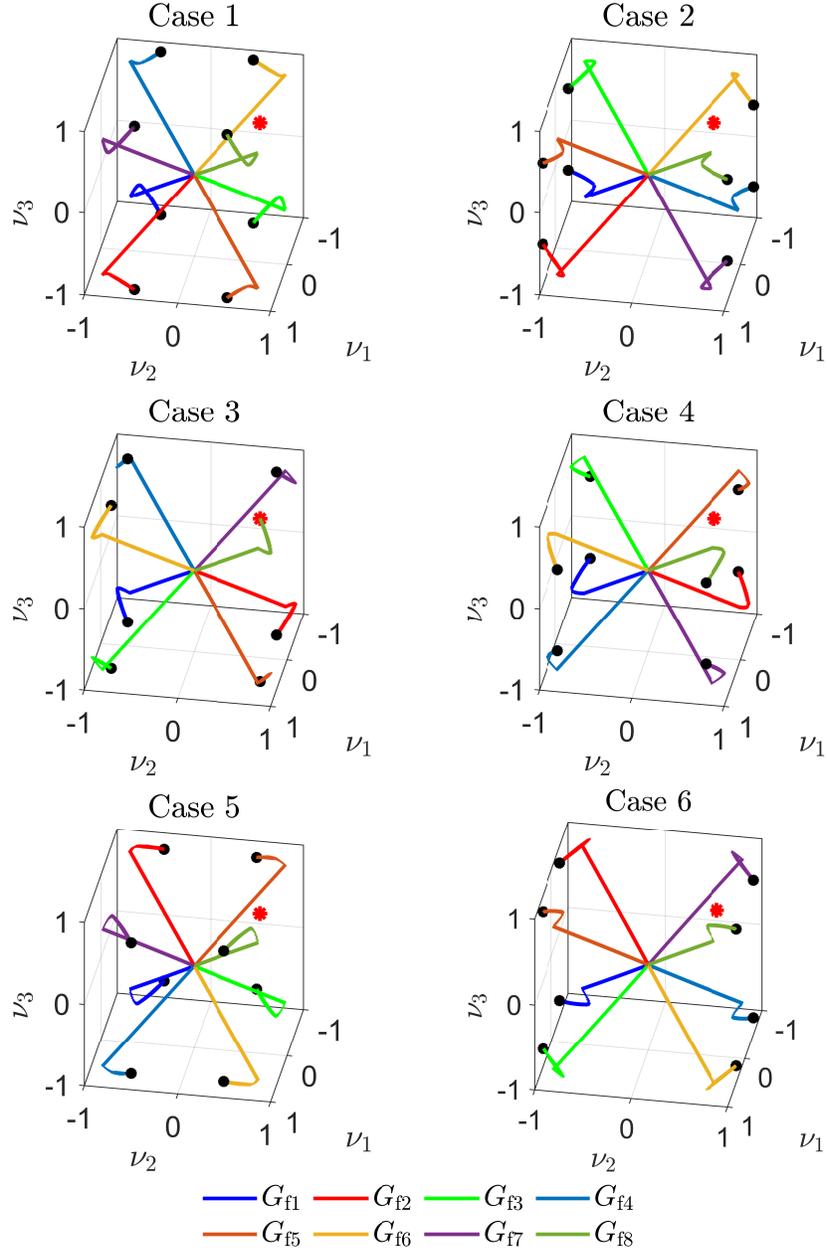}
        \caption 
            {
        	    Estimation of three unknown parameters in the nonlinear system \eqref{eq:NL_NLP_state_lu3}, \eqref{eq:NL_NLP_output_lu3}.
        	    Each case shows the parameter pre-estimate $\nu_{k}$ for the choice of $G_\rmf$ and $\SO_\rmp$ given in  Table \ref{tab:filter_coefficients} and the scalers $\sigma_i$ given in Table \ref{tab:filter_coefficients_sign}.
        	}
        \label{fig:DDDAS_Book_RCPE_NLS_NLP_lu3_all}
    \end{figure}

    For each choice of the scalar multipliers $\sigma_i$, the pre-estimate $\nu_k$ converges to a point in $\SSS_{\SO_\rmp}$ as shown in Figure \ref{fig:DDDAS_Book_RCPE_NLS_NLP_lu3_all}.
    Note that each point in $\SSS_{\SO_\rmp}$ maps to $\mu$, and the parameter estimate $\hat \mu_k$ converges to $\mu$ in each case. 
    
    This example shows that there exists at least one permutation matrix that yields convergence for every choice of the filter coefficients considered in this example. 
    Furthermore, the element of $\SSS_{\SO_\rmp}$ to which the parameter pre-estimate converges depends on $\sigma_i.$
    However, since each element of $\SSS_{\SO_\rmp}$ is mapped to the true value of the unknown parameter, the choice of $\sigma_i$ is not critical. 
    In fact, all possible combinations of $\sigma_i$ yield parameter convergence. 
    \EndExample

    \begin{table}[ht]
        \centering
        \setlength{\tabcolsep}{8pt}
        \subfloat[]{
        \begin{tabular}{|c|c|c|c|c|}
            \hline
            Case          & $N_1   $    & $N_2  $  & $N_3$ &  $\rmp$
            \\ \hline
            1    & $[1 \ 0 \ 0]$   & $[0 \ 1 \ 0]$  & $[0 \ 0 \ 1]$  & $(2,1,3)$
            \\ \hline
            2    & $[1 \ 0 \ 0]$   & $[0 \ 0 \ 1]$  & $[0 \ 1 \ 0]$ & $(3,1,2)$  
            \\ \hline
            3    & $[0 \ 1 \ 0]$   & $[1 \ 0 \ 0]$  & $[0 \ 0 \ 1]$ & $(1,2,3)$
            \\ \hline
            4    & $[0 \ 1 \ 0]$   & $[0 \ 0 \ 1]$  & $[1 \ 0 \ 0]$   & $(3,2,1)$
            \\ \hline
            5    & $[0 \ 0 \ 1]$   & $[0 \ 1 \ 0]$  & $[1 \ 0 \ 0]$   & $(2,3,1)$
            \\ \hline
            6    & $[0 \ 0 \ 1]$   & $[1 \ 0 \ 0]$  & $[0 \ 1 \ 0]$   & $(1,3,2)$
            \\ \hline
        \end{tabular}
        \label{tab:filter_coefficients}
        }
        \quad 
        \subfloat[]{
        \begin{tabular}{|c|r|r|r|}
            \hline
            Filter          & $\sigma_1   $    & $\sigma_2  $  & $\sigma_3$ 
            \\ \hline
            $G_{\rmf 1}$    & $1$   & $1$  & $1$
            \\ \hline
            $G_{\rmf 2}$    & $-1$   & $1$  & $1$
            \\ \hline
            $G_{\rmf 3}$    & $1$   & $-1$  & $1$
            \\ \hline
            $G_{\rmf 4}$    & $1$   & $1$  & $-1$
            \\ \hline
            $G_{\rmf 5}$    & $-1$   & $-1$  & $1$
            \\ \hline
            $G_{\rmf 6}$    & $1$   & $-1$  & $-1$
            \\ \hline
            $G_{\rmf 7}$    & $-1$   & $1$  & $-1$
            \\ \hline
            $G_{\rmf 8}$    & $-1$   & $-1$  & $-1$
            \\ \hline
        \end{tabular}
        \label{tab:filter_coefficients_sign}
        }
        \caption
            {
                Filter coefficients and the permutation used in RCPE to estimate the unknown parameters in the nonlinear system  \eqref{eq:NL_NLP_state_lu3}, \eqref{eq:NL_NLP_output_lu3}.  
            }
        \label{tab:NonLin_lu_3_Ex3}
    \end{table}

\clearpage
\section{ Parameter Estimation in the Generalized Burgers Equation}
\label{sec:Burgers}
 
Consider the generalized one-dimensional viscous Burgers equation \cite{blackstock1985generalized}
\begin{align}
	\frac{\partial u}{\partial t} +
		  	\mu_1 \frac{\partial }{\partial x} \frac{u^2}{2} 
        =
            \frac{\partial }{\partial x} 
            \left( \mu_2 \frac{\partial u}{\partial x} \right),
	\label{eq:BurgersEquation}
\end{align}
where $u(x,t) $ is a function of space and time with domain $[0,1]\times [0,\infty)$, $\mu_1>0$ is the convective constant, and $\mu_2>0$ is the viscosity.
Note that there is no external input to the system \eqref{eq:BurgersEquation} and $u$ denotes the solution of this partial differential equation.
Let the initial condition be $u(x,0)=0$ for all $x\in[0,1]$, and consider the boundary conditions $u(0,t)=0$ and $u(1,t) = \sin (5t)+0.25\sin(10t)$ for all $t\ge0.$ 
The Burgers equation \eqref{eq:BurgersEquation} is discretized using a forward Euler approximation for the time derivative, a second-order-accurate upwind method for the convective term, and a second-order-accurate central difference scheme for the viscous term. 
The spatial domain $[0,1]$ is discretized using $N$ equally spaced grid points;
thus $\Delta x \isdef \frac{1}{N-1}$.
The time step $\Delta t$ is chosen to satisfy the Courant–Friedrichs–Lewy (CFL) condition, that is, 
\begin{align}
    \Delta t 
        <
            \frac{C_{\rm max} \Delta x}{| {\rm max}(u) |},
\end{align}
where the Courant number $C_{\rm max}$ depends on the discretization scheme \cite{courant1967partial}. 
Finally, the discrete variable $u_{j,k} \isdef u((j-1) \Delta x, k \Delta t)$ is defined on the grid points $j \in \{1,\ldots, N\}$ for all time steps $k \ge0$.
Hence, at each grid point, $j \in \{3,\ldots, N-1\}$, 
\begin{align}
	u_{j,k+1} 
	    = 
		    u_{j,k} &- 
		    \mu_1
		    \frac{\Delta t}{2 \Delta x} 
		    (1.5 u_{j,k}^2 - 2 u_{j-1,k}^2 + 0.5 u_{j-2,k}^2) + 
		    \nn \\
		    &\quad
		    \mu_2 \frac{\Delta t}{\Delta x^2} 
		    (u_{j+1,k} - 2u_{j,k} + u_{j-1,k}).
	\label{eq:disc_BurgersEquation}
\end{align}
For all $k\ge0,$ the discretized boundary conditions are
\begin{align}
    u_{1,k} = u_{2,k} = 0,\quad 
    u_{N,k}
        = 
            \sin (5\Delta t k) + 0.25 \sin (10\Delta t k)  ,
    \label{eq:disc_BE_BC}
\end{align}
and, for all $j\in\{3,\ldots,N-1\},$ the initial condition is
\begin{align}
    u_j(0)
        = 
            0.
    \label{eq:disc_BE_IC}
\end{align}
Finally, for all $k\ge0,$ let the measurement $y_k$ be given by
\begin{align}
    y_k 
        \isdef 
            u_{87,k}
        =
            u(0.87, k \Delta t).
\end{align}

\begin{figure}[ht]
	\centering
	 \subfloat[]
	    {
	        \includegraphics[height=0.4\columnwidth]{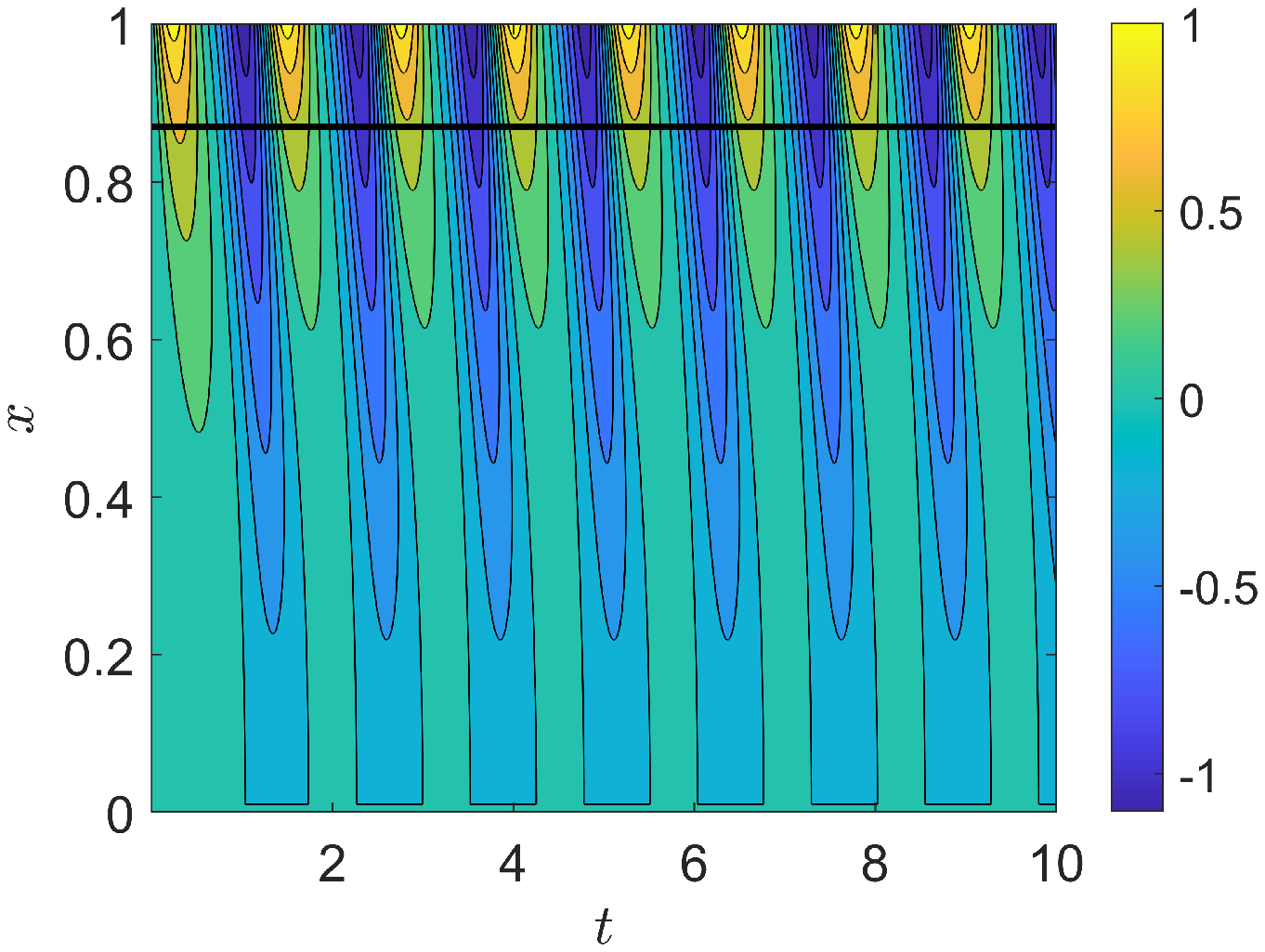}
	        \label{fig:DDDAS_Book_RCPE_Burgers_lu2_Sample}
        }
    \subfloat[]
	    {
	        \includegraphics[height=0.35\columnwidth]{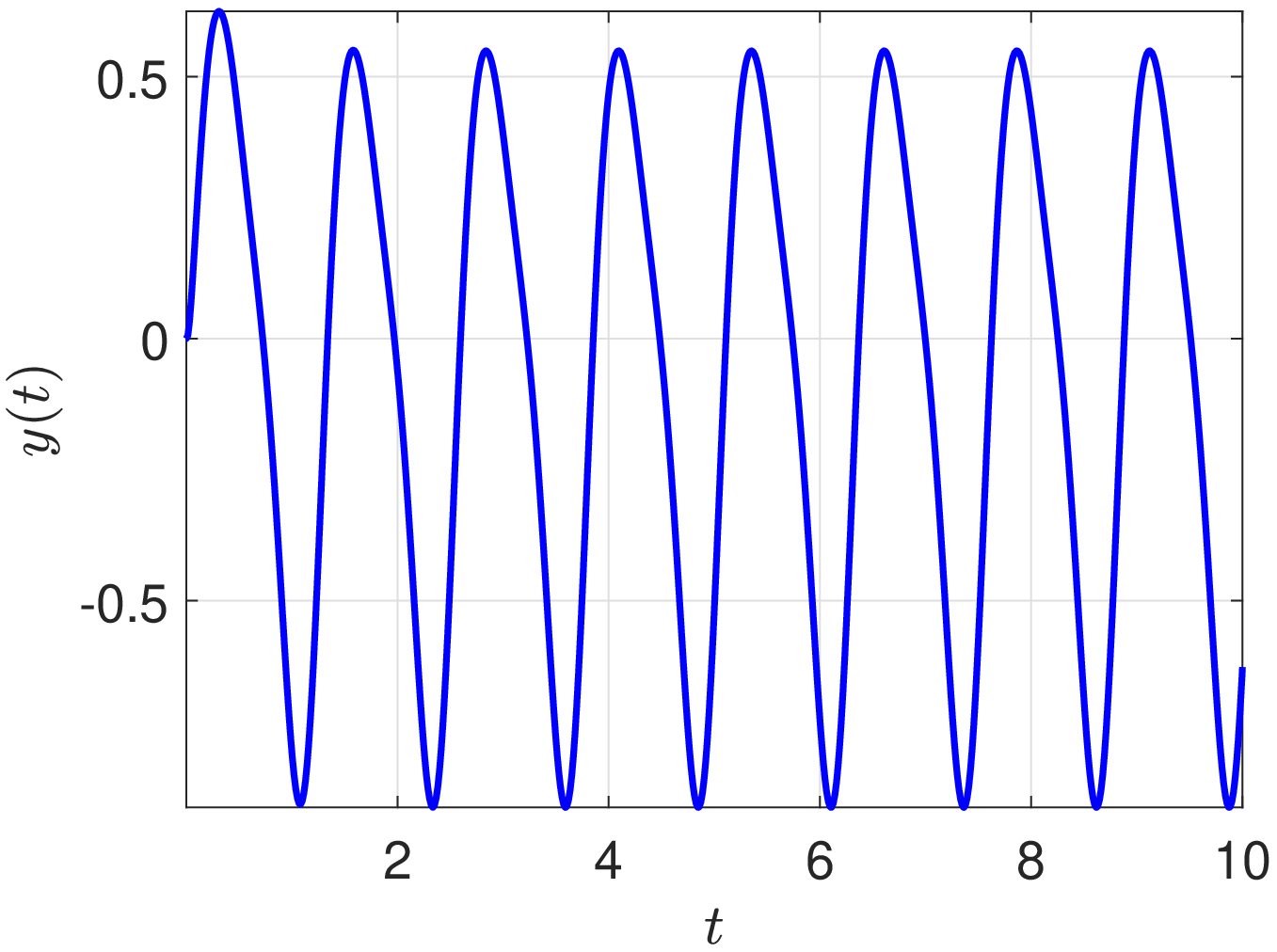}
	        \label{fig:DDDAS_Book_RCPE_Burgers_lu2_Sample_y}
        }
    \caption 
    	{
    	   Simulation of the generalized Burgers equation \eqref{eq:BurgersEquation} with the discretization \eqref{eq:disc_BurgersEquation}.
    	   (a) shows the variable $u(x,t)$, and
    	   (b) shows the measurement $y(t)$.
    	   The magnitude of $u(x,t)$ is shown in the color bar, and the black line denotes the location of the measurement.
    	}
    \label{fig:DDDAS_Book_RCPE_Burgers_lu2_Sample_1}
\end{figure}

In this example, $\mu_1 = 1.4, \mu_2 = 0.3$, $C_{\rm max} = 0.25$, $N=100$, and $\Delta t = 10^{-4}$ s. 
Figure \ref{fig:DDDAS_Book_RCPE_Burgers_lu2_Sample} shows the numerical solution of \eqref{eq:disc_BurgersEquation} with the boundary conditions \eqref{eq:disc_BE_BC} and initial conditions \eqref{eq:disc_BE_IC}, where the solid black line shows the measurement location. 
Figure \ref{fig:DDDAS_Book_RCPE_Burgers_lu2_Sample_y} shows the measurement $y_{k}.$
%
The objective is to estimate the unknown parameter $\mu \isdef [\mu_1\ \ \mu_1]^\rmT$ using the scalar measurement $y_k$.
Thus, $l_\mu=2$ and $l_y=1.$

\begin{figure}[ht]
	\centering
	\includegraphics[width=01.0 \columnwidth]{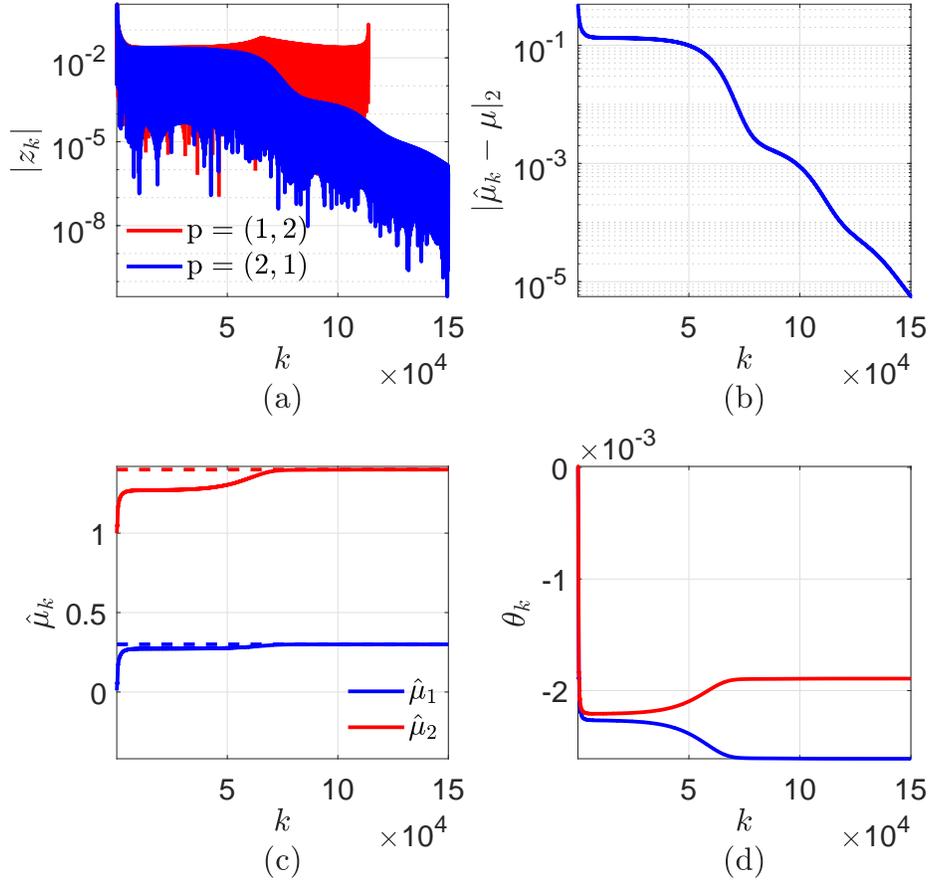}
    \caption 
        [Estimation of unknown parameters for the generalized Burgers equation. ]
    	{
    	    Estimation of two parameters in the generalized Burgers equation \eqref{eq:BurgersEquation}.
    	    (a) shows the output error for both permutation matrices, 
    	    (b) shows the parameter error, 
    	    (c) shows the parameter estimates, and
    	    (d) shows the parameter estimator coefficients for $\rmp=(2,1).$
    	}
    \label{fig:DDDAS_Book_RCPE_Burgers_lu2}
\end{figure}

In order to start the estimation model, nonzero values of $\hat \mu_1(0)$ and $\hat \mu_2(0)$ are needed.
A simple way to ensure nonzero values of the estimates is to replace $\mu$ by $\hat \mu_{k} = \overline \mu + \SO_\rmp |\nu_{k}|$, where $\overline \mu = [\overline\mu_1\ \ \overline\mu_1]^\rmT =  [1 \ \ 0.01]^\rmT$, so that $\hat \mu(0) \ne 0$ .
In RCPE, let $N_1 = [1 \ 0]$, $N_2 = [0 \ 1]$, $\lambda = 0.9999,$ and $R_\theta = 10^6 I_2$. 
Let $\rmp = (2,1)$ so that $\SSS_{\SO_\rmp} = \{ [\pm(\mu_2-\overline\mu_2) \ \pm(\mu_1-\overline\mu_2)  ]^\rmT \}$.
Figure \ref{fig:DDDAS_Book_RCPE_Burgers_lu2} shows the output error, norm of the parameter error, the parameter estimates, and estimator coefficient.
Figure \ref{fig:DDDAS_Book_RCPE_Burgers_lu2} also shows the diverging output error for the permutation $\rmp=(1,2)$.
%

This example shows that RCPE can successfully estimate unknown parameters in large-scale computational models without using ensembles and gradients. 
UKF, on the other hand, would have required an ensemble of 203 models each with 101 state updates resulting in over 20,000 state updates at each step. 
Furthermore, the sigma points constructed by UKF to propagate the covariance may violate the physical laws or may excite unstable modes in the dynamics, and thus some ensemble members may diverge.
In contrast, RCPE updates only the parameter estimate, which can be saturated to constrain the estimate to physically realistic parameter values. 
\EndExample

 \clearpage
\section{Global Ionosphere-Thermosphere Model}
\label{sec:GITM}
GITM is a computational code that models the thermosphere and the ionosphere of the Earth as well as that of various planets and moons by solving coupled continuity, momentum, and energy equations \cite{ridley2006global}. 
By propagating the governing equations, GITM computes neutral, ion, and electron temperatures, neutral-wind and plasma velocities, and mass and number densities of neutrals, ions, and electrons. 

GITM uses a uniform grid in latitude with width $\frac{2 \pi }{n_{\rm{lat}}}$ rad, where $n_{\rm{lat}}$ is the number of grid points.
In longitude and altitude, GITM uses a stretched grid to account for temperature and density variations. 
GITM is implemented in parallel, where the computational domain (the atmosphere from 100 km to 600 km) is divided into blocks.
Ghost cells border the physical blocks to exchange information. 
GITM can be run either in  one-dimensional mode, where horizontal transport is ignored, or in global three-dimensional mode. 
Furthermore, GITM can be run at either a constant or a variable time step, which is calculated by GITM based on the physical state and the user-defined CFL number in order to maintain numerical stability. 

To initialize GITM, neutral and ion densities and temperatures for a chosen time are set using the Mass Spectrometer and Incoherent Scatter radar (MSIS) model \cite{hedin1991extension} and International Reference Ionosphere (IRI) \cite{bilitza2001international}.
The model inputs for GITM are $10.7$ cm solar radio flux (F10.7 index), hemispheric power index (HPI), interplanetary magnetic field (IMF), solar wind plasma (SWP), and solar irradiance, all of which are read from a text file containing the time of measurements and the measured values.
%
These signals are available from various terrestrial sensor platforms.
%

RCPE has previously been used to compensate for the dynamic cooling process \cite{tonydatamining}, 
estimate the solar irradiance F10.7 \cite{ali2015retrospective}, 
and estimate the eddy diffusion coefficient \cite{ankit_EDC_ACC2018}
in GITM.
In this study, parameters modeling the thermal conductivity coefficient that model the thermal conductity in GITM are estimated. 

The temperature at a given altitude depends on the vertical temperature gradient is described by
\begin{align}
    \frac{\delta T}{\delta t} \propto \frac{\delta }{\delta z} \lambda_\rmc  \frac{\delta T}{\delta z},
\end{align}
where $T$ is  absolute temperature, $t$ is time, and $z$ is  altitude.
The thermal conductivity  $\lambda_\rmc$ is modeled as
\begin{align}
    \lambda_\rmc = \sum_{i \in\{ {\rm O, O_2, N_2}\}} \left(
        \frac{N_i}{N_{\rm total}}
    \right)
    \kappa_{i} T^S,\label{TCandSeqn}
\end{align}
where
$\kappa_{i}$ is the thermal-conductivity coefficient of the $i$th species,
$N_{i}$ is the number density  of the $i$th species, and
$S$ is the temperature exponent. 
The thermal conductivity parameterizes the hydrodynamic transport process that describes the movement of heat through the atmosphere. 
In GITM, heat is transferred by collisions of the particles in the thermosphere (100 km to 600 km).
The three main constituents in the thermosphere are molecular nitrogen $\rm N_2$ and oxygen $\rm O_2$ and atomic oxygen $\rmO$, with the molecular species dominating below about 200 km altitude. Therefore, each of these constituents are included in \eqref{TCandSeqn}.
Furthermore, since the molecular species conduct heat in similar ways, the thermal-conductivity coefficient $\kappa_{\rmN_2}$ is assumed to be equal to the thermal-conductivity coefficient $\kappa_{\rmO_2}$ in \eqref{TCandSeqn}.
In this example, the unknown parameters $\kappa_{\rmO}, \kappa_{\rmO_2}, $ and $S$ in \eqref{TCandSeqn} are estimated.

\begin{figure}[ht]
	\centering
	\includegraphics[width=01.0\columnwidth]
	{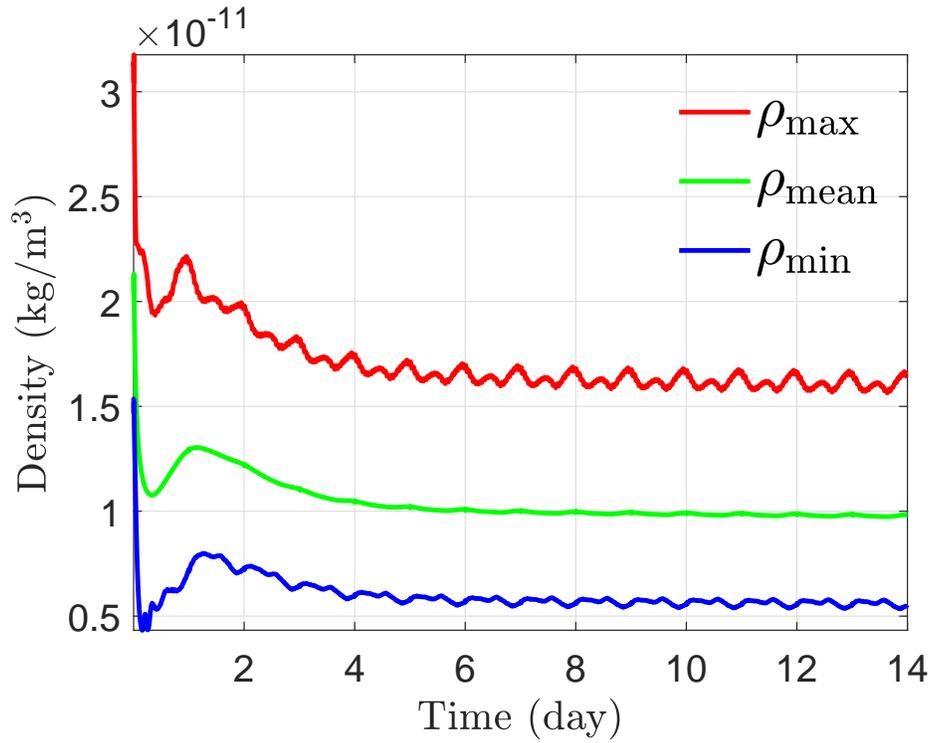}
    \caption 
    	{
    	    Simulated neutral-density measurements. These measurements are synthesized using the computed neutral density at the altitude 300 km.
    	}
    \label{fig:GITM_TC_True_densities}
\end{figure}

To obtain neutral-density data, the thermosphere is simulated using GITM from 11-21-2002 to 12-05-2002 with the constant thermal-conductivity coefficients $\kappa_{\rmO_2} = 3.6 \times 10^{-4} \ {\rm m^2/s}$ and $\kappa_\rmO = 5.6 \times 10^{-4} \ {\rm m^2/s}$ as well as the  temperature exponent $S = 6.9 \times 10^{-1}$.
The neutral density is sampled at intervals of 1 min, and thus the simulated time, which is measured in days, is given by $k/1440,$ where $k\ge0$ is the number of simulated minutes.
At each sample time, the global minimum density  $\rho_{\rm min}$, the global maximum density $\rho_{\rm max},$ and the mean density  $\rho_{\rm mean}$ at the altitude 300 km are computed. 
This \redtext{simulated} data, shown in Figure \ref{fig:GITM_TC_True_densities}, is used by RCPE in the examples below.
Only the data after 2 hr (0.083 day) is used in order to avoid the initial transient.

\begin{figure}[ht]
    	\centering
    	\includegraphics[width=01.0\columnwidth, trim= .5cm 0  1.6cm 0, clip]
    	{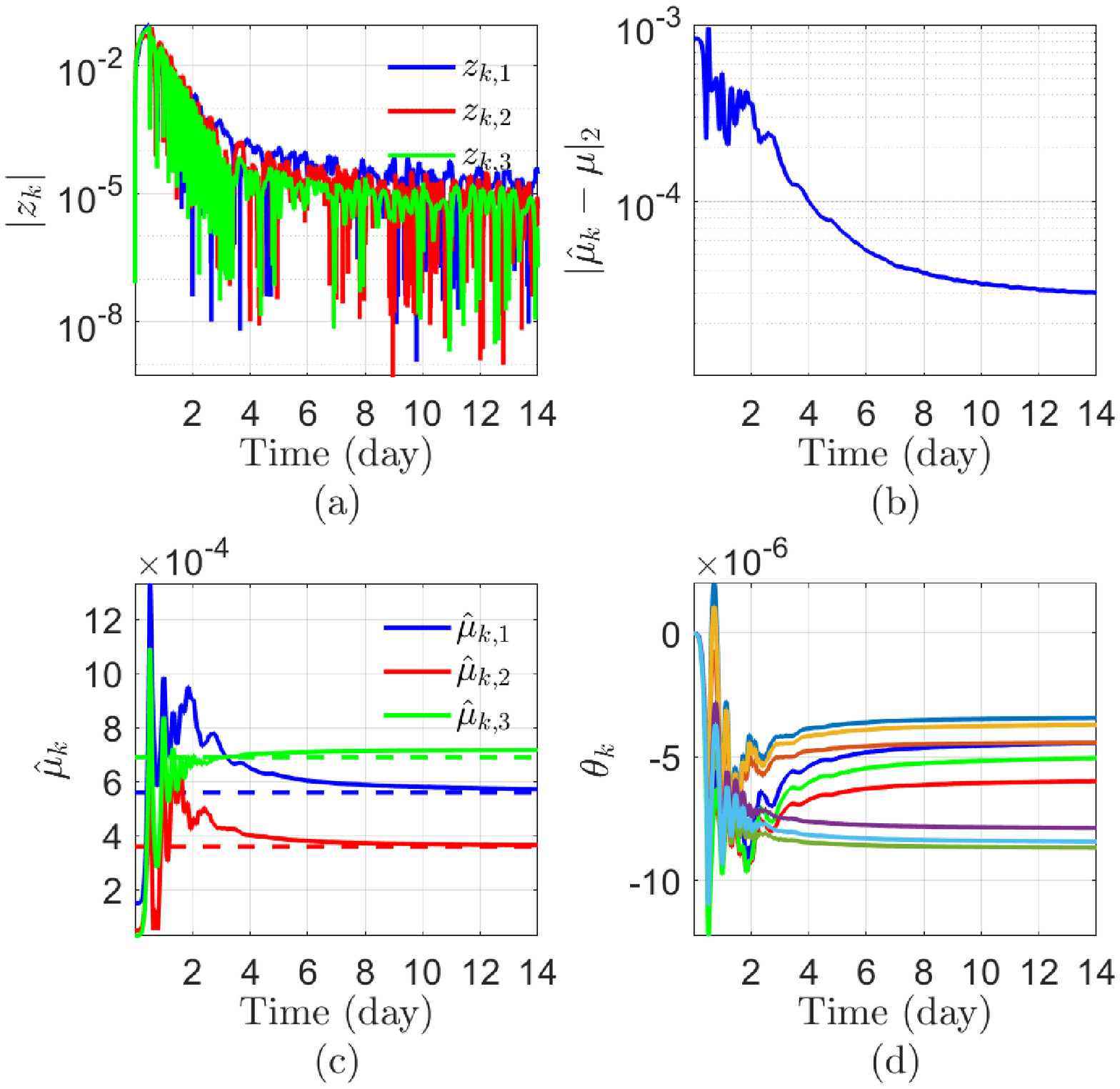}
        \caption 
        	{
        	    Estimation of $\kappa_{\rmO_2},\kappa_{\rmO},$ and $S$ using  measurements of $\rho_{\rm min}, \rho_{\rm max}$, and $\rho_{\rm mean}$.
        	    (a) shows the output error, 
        	    (b) shows the parameter error, 
        	    (c) shows the parameter estimates, and
        	    (d) shows the parameter estimator coefficients.
        	}
        \label{fig:DDDAS_chapter_GITM_lmu3_lz3_VRF_44}
    \end{figure}

\begin{figure}[ht]
    	\centering
    	\includegraphics[width=01.0\columnwidth]
    	{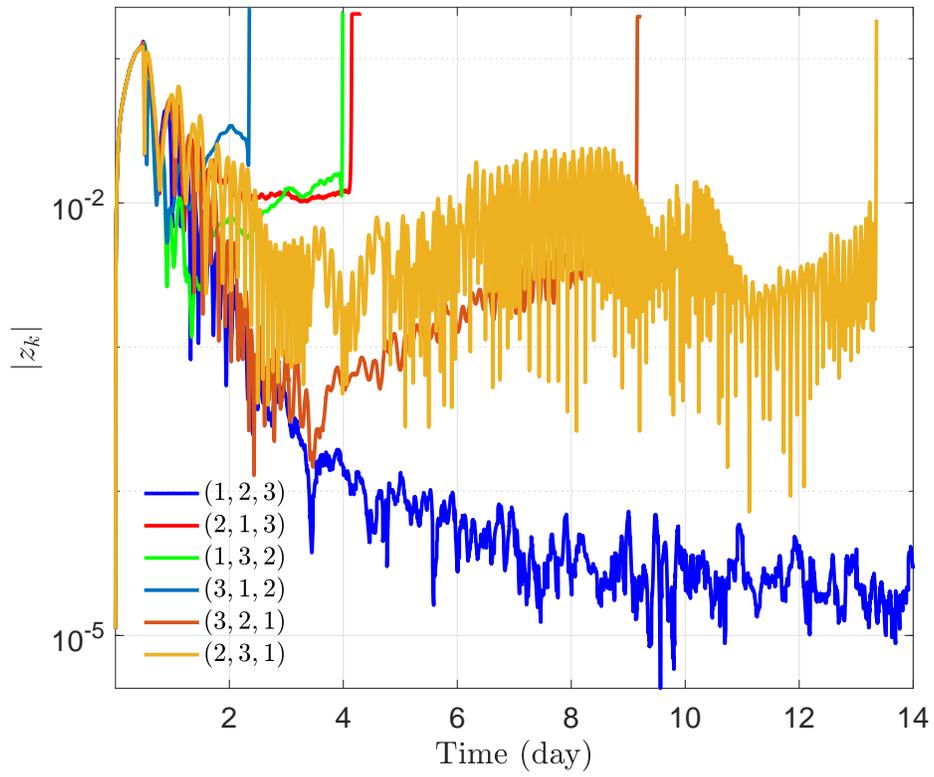}
        \caption 
        	{
        	    Estimation of $\kappa_{\rmO_2},$ $\kappa_{\rmO},$ and $S$ using  measurements of $\rho_{\rm min},$ $\rho_{\rm max}$, and $\rho_{\rm mean}$.
        	    The output error is shown  for all six permutations of the estimates of $\kappa_{\rmO_2}$, $\kappa_{\rmO}$, and $S$ using $\rho_{\rm min}$, $\rho_{\rm max}$, and $\rho_{\rm mean}$.
        	    Note that the output error converges for the permutation (1,2,3), but diverges for the remaining five permutations.
        	}
        \label{fig:GITM_lmu3_lz3_all_perms}
    \end{figure}

In the estimation model, the thermal conductivity coefficients estimates are given by
    \begin{equation}
        \matl{c}
            \hat \kappa_{\rmO_2,k} \\
            \hat \kappa_{\rmO,k}  \\
            \hat S_k
        \matr
            =
                \matl{c}
                    \hat \kappa_{\rmO_2,0} \\
                    \hat \kappa_{\rmO,0} \\
                    \hat S_0
                \matr
                + 
                \SO_\rmp \vert M \nu_k \vert,
    \end{equation}
    where
    $\hat \kappa_{\rmO_2,0}= 1.5 \times 10^{-4}\, \rm m^2/s$ is the initial estimate of $\kappa_{\rmO_2}=0.5 \times 10^{-4}\, \rm m^2/s$, 
    $\hat \kappa_{\rmO,0}$ is the initial estimate of $\kappa_{\rmO}$,
    $\hat S_{0}=3 \times 10^{-1}$ is the initial estimate of $S$,
    $\SO_\rmp=I_3$,
    $M={\rm diag}(1,1,1000)$,
    and
    $\nu_k$ is given by \eqref{eq:nu_regressor_form}.
    Note that the scaling matrix $M$ ensures that all elements of $\nu_k$ have similar magnitude.
    The output of the estimation model is 
    $\hat y_k = 10^{10}[
    \hat \rho_{{\rm min}, k} \ \
    \hat \rho_{{\rm mean}, k} \ \
    \hat \rho_{{\rm max}, k}
    ]^\rmT$, where the scaling ensures that the error $z_k$ and the parameter estimate $\hat \mu_k$ have similar magnitude. 
    In RCPE, 
    Figure \ref{fig:DDDAS_chapter_GITM_lmu3_lz3_VRF_44} shows the output error, the estimates of $\kappa_{\rmO_2}$, $\kappa_{\rmO}$, and $S$, the norm of the parameter error $\hat \mu_k-\mu$, and  the parameter estimator coefficients $\theta_k$.
    %
    Figure \ref{fig:GITM_lmu3_lz3_all_perms} shows the output error for all six choices of $\rmp$. 
    Note that the output error diverges for all but one choice of $\rmp$.

    This application shows that RCPE is an effective technique for removing various sources of model bias using a limited set of observational data.
    Additionally, the density estimates are also improved as the parameter estimate is improved by RCPE.  
    \EndExample


\clearpage
\section{Conclusions}
\label{sec:conclusions}

Motivated by the need to estimate representational parameters in large-scale models in \redtext{DDDAS systems}, this chapter reviewed the retrospective cost parameter estimation technique. 
This approach to parameter estimation is both modular and computationally inexpensive since it does not rely on the computation of gradients or the propagation of ensembles. 
However,  in the worst case scenario, the RCPE algorithm may need to be executed $l_\mu!$ times in order to determine the parameter permutation that yields convergence of the parameters to the true values. 
%

\section*{Acknowledgments}

The authors thank Aaron Ridley for help with GITM and Karthik Duraisamy for assistance with the Burgers equation.


This research was supported by AFOSR under DDDAS grant FA9550-16-1-0071
(Dynamic Data-Driven Applications Systems \href{http://www.1dddas.org/}{http://www.1dddas.org/} ) 
and NSF grant CNS-0539053 under the NSF DDDAS Program.


\printbibliography

\end{document}